\documentstyle[12pt,aps,preprint]{revtex}
\tightenlines
\newcommand{\ups}\Upsilon

\begin{document}

\title{On The Invisible Decays of the $\Upsilon$ and $J/\Psi$ Resonances }
\author{L.N. Chang,\thanks{laynam@vt.edu} 
O. Lebedev\thanks{lebedev@quasar.phys.vt.edu}}
\address{\em Virginia Polytechnic Insitute and State University\\
Department of Physics\\
Institute for Particle Physics and Astrophysics\\
%\bigskip
%\bigskip
Blacksburg, Virginia 24061}
\author{J.N. Ng\thanks{misery@triumf.ca}}
\address{\em TRIUMF, 4004 Wesbrook Mall, Vancouver, B.C. Canada V6T 2A3 }
\date{ Nov 3, 1997}  %February 9/98 JT
\maketitle
\thispagestyle{empty}

\begin{abstract}
We estimate the most important corrections to the branching ratios for
the invisible decays of quarkonium states, arising from possible
extensions
of the Standard Model.  Among the possibilities considered are the
presence
of extra $Z$-bosons, minimal supersymmetric extensions of the Standard
Model
with R-parity violation and decays into Goldstinos.  Prospects of
detecting these corrections
at existing and future B-factories and $\tau$-charm factories are
discussed.
\end{abstract}

%\baselineskip=24pt plus 2pt minus 1pt
%\begin{center}
%(submitted to .....)
%\end{center}

\newpage

\hspace*{\parindent}
B-meson factories under construction at KEK and at SLAC [1] can deliver
$10^{8}$
$\ups$'s and $\ups^{\prime}$'s per year, when they are tuned to run at
resonance.  Coupled with what is available at CLEO III, these facilities
will
make it possible
for the first time to study in detail invisible decays of these
resonances.
Such decay modes can be studied by observing the decay
$\ups^{\prime} \to \ups + 2 \pi{\rm 's}$ at resonance and tagging on the
invariant
mass of the dipion system at the $\ups$-mass.

Invisible decays of heavy quarkonium states offer a window into what may
lie beyond the Standard Model.  The reason is that apart from neutrinos,
the minimal Standard Model predicts no other channels that these states can
decay into.  The associated rates for these decays can be computed
precisely, and so any observed departure can furnish hints of
structures over and above those in the Standard Model.  As we will show
below, a more complete test of what these structures might be would
require
similar measurements on the $J/\Psi$-invisible decay widths, something
which
could be achieved at the $\tau$-charm factory in the future.

In what follows, we concentrate on what we believe to be
the largest among  these effects.
In particular, we will estimate the branching ratios into neutrinos in the
presence of extra $Z$-bosons, and R-parity violating effects in
supersymmetrized
Standard Models, and finally decays into Goldstinos.

To begin, we present the Standard Model prediction
for the branching ratio of the invisible decays of
$\Upsilon$ and $J/ \Psi $ and their observed decays into
electron-positron pairs. To our knowledge the detail formulae have not
been
given before.
Within the  Standard Model the invisible mode consists solely of decays
into
three types of neutrino-antineutrino pairs.
Neglecting polarization effects and taking
into account $e^{+} e^{-} $ production through a photon only we get
\begin{eqnarray}
 \frac{\Gamma \;( \Upsilon \rightarrow \nu \bar \nu) }
         {\Gamma \;( \Upsilon \rightarrow e^{+} e^{-} )} &=&
         \frac {27 G^2 M^4_{\Upsilon} }{64 \pi^2 \alpha^2}
         (-1 + \frac{4}{3} \sin^2 \theta_{W} )^2   \; \\
           &=& 4.14\times 10^{-4}\;, \nonumber  \\
  \frac{\Gamma (J/ \Psi \rightarrow \nu  \bar \nu) )}
         {\Gamma (J/ \Psi \rightarrow e^{+} e^{-} )} &=&
         \frac {27 G^2 M^4_{J/ \Psi} }{256 \pi^2 \alpha^2}
         (1 - \frac{8}{3} \sin^2 \theta_{W} )^2   \;, \\
      &=& 4.54 \times 10^{-7} \;, \nonumber
\end{eqnarray}
with $G$ and $\alpha$ being the Fermi and the fine structure constants
respectively.
$M_{\Upsilon , J/ \Psi}$ are masses for
the $\Upsilon$ and $J/
\Psi$ states.
 These formulas are expected to
be correct up to about 2-3\%.
The major sources
of theoretical uncertainty involved in (1)
and
(2) can be listed as follows:

\begin{enumerate}
%1
\item corrections to the $\Upsilon$ and  $J/ \Psi$ wave functions,
including
QCD corrections and polarization effects,
%2
\item  $e^{+} e^{-} $ production through Z,
%3
\item  electroweak radiative corrections,
%4
\item  corrections due to the Higgs boson.
\end{enumerate}

The first correction is not expected to be significant since it similarly
modifies
the neutrino and electron decay widths, and so the leading order cancels
in the branching ratio.
The second correction will introduce a factor $0.1M_{Q}^2/M_Z^2$, and so
is
not expected to be significant either.

The third correction can be substantial, and thus
needs to be considered in a little more detail.
In the numerical calculation quoted above, we have
included the ${\overline{MS}}$-running of the coupling constants in
$\alpha$,
$G$, as
well as in
$\sin^{2}\theta$.
We have chosen the $Z$-mass,  $G$, and
$\alpha(M_Z)$ as the input parameters and run the
other parameters from their values at the $Z$-mass down to the low
energies
where the quarkonia reside.
These are by far the most significant
contributions.  To the accuracy we are working in, of the
order of 2-3\%, we may ignore threshold effects.
Box-type diagrams with double Z ($W^{+},W^{-}$) or photon emission
which
may be expected to modify the tree level results turn out to be
negligible. The largest of these
 with two internal photon lines  vanish due to  charge- conjugation
symmetry since both $\Upsilon$ and  $J/ \Psi$ are C-odd.
The contribution of the
remaining box diagram with
virtual $Z$ and $\gamma$ also makes only a negligible correction to the
tree
level
Z-mediated $e^{+} e^{-} $-production and can be discarded as well.
Finally, the Higgs correction is vanishingly small simply because we are
dealing with virtually massless particles in the final state.

%%30 Nov, 1997-7 pm%%
As a result,
branching ratios of  (1) and (2) are theoretically clean and thus offer
a rare opportunity to search for physics beyond the
Standard Model if they are relatively large.
Among the most viable
candidates are
supersymmetry (SUSY) and Grand Unified Theories (GUTs).
In the case of spontaneously broken SUSY the decays of $\Upsilon$ and
$J/ \Psi$ will produce invisible Goldstinos (or light gravitinos) in
addition to the
neutrino models. All other particles in the final state are prohibited by
kinematical considerations (see [2]). Therefore,
measuring the
invisible width and comparing it with the SM prediction could provide
information about new physics. In the case of SUSY models, we can gain
information on the
SUSY-breaking scale as well as about the existence of R-parity violating
terms in the superpotential.

We first consider the case of decay into
light gravitinos denoted by $\tilde{g}$. The general structure of the
Goldstino interactions is known,
and given for example in [3].

$\Upsilon$ and
$J/ \Psi$ can decay into Goldstinos via virtual $Z,\gamma$ in the
s-channel and
 via exchanges of $b,c$-squarks in the t-channel. Neglecting the
 Goldstino exchange effects, the corresponding
rates for the $J/\Psi$ are calculated to be
\begin{eqnarray}
&& \frac{\Gamma \;( J/ \Psi \buildrel Z \over \rightarrow
\tilde{g}\overline{\tilde{g}})}
{\Gamma (J/ \Psi \rightarrow e^{+} e^{-} )}
=
\frac{9 G^2 M_{J/ \Psi}^8 v^4 \; \cos^2 2\beta}{4096 \pi^2 \alpha^2 F^8}
(1 - \frac{8}{3} \sin^2 \theta_{W} )^2  ,\\
&& \frac{\Gamma \;( J/ \Psi \buildrel {\rm c- squark} \over
\longrightarrow
\tilde{g}\overline{\tilde{g}})}
{\Gamma (J/ \Psi \rightarrow e^{+} e^{-} )}
= \frac{9 M^2_{J/\Psi} m_{c}^{10}}{32 \pi^2 \alpha^2
m_{\tilde c }^4 F^8} \;,
\end{eqnarray}
where $v^2=v_1^2+v_2^2\approx  (174\;GeV)^2$, $\;$ v$_{1,2}=\langle
\Phi_{1,2}^0 \rangle$,
$\; \tan \beta =v_2 / v_1 $, $m_c(\tilde{m}_c )$ is the mass of the
c-quark
(squark) and $\;F$ is the SUSY-breaking scale. The photon channel is
suppressed as compared to the $Z$ and $\tilde c$ ones since the
photon-goldstino coupling containes higher powers of $F$ and can be
neglected in the leading order approximation.
  Unfortunately, for a reasonable
choice for the value of $F\sim 1$~TeV[3],
    these rates
are extremely small and far beyond the experimental capabilities.
Doing the same calculation for the $\Upsilon$ does not improve matters
for gravitino decay modes. The above considerations rule out light
gravitinos as candidates for the invisible decays.

Nevertheless, supersymmetry still can affect the invisible width because
of
R-parity breaking processes.  Such processes affect neutrino decay modes
through
squark exchange. In terms of superfields, the relevant interactions are
generated by a
$\lambda '_{ijk} L^{i}_{L} Q^{j}_{L} \bar D^{k}_{R}$  term in
the
superpotential.  Here $i,j,k$ are the generation indices and we have
suppressed SU(2) indices [4].
Expressed in terms of component fields the interaction Lagrangian takes
the form
\begin{eqnarray}
-{\cal L}_{\not{R}} &=& \lambda'_{ijk} [\tilde{\nu}_{iL}
\overline{d}_{kR}
d_{jL} + \tilde{d}_{jL} \overline{d}_{kR} \nu_{iL} + \tilde{d}^*_{kR}
\overline{\nu}^c_{iL} d_{jL} \nonumber \\
& & \mbox{} - (\tilde{e}_{iL} \overline{d}_{kR} u_{jL} + \tilde{u}_{jL}
\overline{d}_{kR} e_{iL} + \tilde{d}^*_{kR} (\overline{e}^c_{iL})
u_{jL})] + {\rm H.c.}
\end{eqnarray}
where we have neglected mixing among generations.  From Eq.~(6) it
can be seen that the terms leading to
neutrino final states involve
the down quarks only, and the $J/ \Psi$ width {\em will not} be affected.
It is noteworthy that one needs to take into account non-SM
corrections to the neutrino widths only, since their contributions
to $e^{+}e^{-}$ production are far less than 0.1\% and lead to
higher order corrections to the branching ratio. Neglecting
possible squark mixings and quark mixings we get
\begin{eqnarray}
&& \frac{\Gamma \;( \Upsilon \rightarrow \nu \bar \nu) }
         {\Gamma \;( \Upsilon \rightarrow e^{+} e^{-} )}
         \Bigg \vert_{\rm SM + SUSY} =
   \frac{9 G^2 M_{\Upsilon}^4}{64 \pi^2 \alpha^2}
   \Bigl( \;
   2 (-1 + \frac{4}{3} \sin^2 \theta_{W} )^2 + \nonumber\\
&& \Bigl[ -1 + \frac{4}{3} \sin^2 \theta_{W} +
   2 \Bigl( \frac{M_{Z}^2}{m_{\tilde b_{R}}^2} +
   \frac{M_{Z}^2}{m_{\tilde b_{L}}^2} \Bigr)
   \;\frac {\cos^2 \theta_{W}}{g^2} \sum_{i=1}^{3}\vert
\lambda'_{i33}\vert^2
   \Bigr]^2
   \Bigr)
\end{eqnarray}
One notices that R-breaking contributions add coherently to the SM result and
reduce the width. The present experimental constraints on
$\lambda_{i33}'$ are very loose [5] and
 SUSY corrections may turn out to be quite significant for this
reaction. For a SUSY mass of 100~GeV, $\lambda'_{133} < 0.002$ from
$\nu_e$
mass calculation, $\lambda'_{233} < 0.4$ and $\lambda'_{333} < 0.26$ from
the ratio of hadronic to leptonic widths at the Z-pole [6].
We note that Eq. (6)  also includes an incoherent piece coming
from
decays into neutrinos of different flavors.

We display in Fig. 1 the branching ratio of Eq.~(1) as a function of the
parameter $x(\rm SUSY)=$ $\displaystyle \sum_{i=1}^{3}$ $\frac{\mid
  \lambda'_{i33}\mid^2}{m^2_{\tilde{b}}}$  where we have set
$m_{\tilde{b}} =
    m_{\tilde{b}_{R}} = m_{\tilde{b}_{L}}=100$GeV/$c^2$ for definiteness. 
It is seen that 
corrections
to the SM result as large as 30\%
are possible for a range of R-parity violating couplings.
We have cut off the $x$ values at the maximally allowed value of
$x_{\rm max} = 2.2 \times 10^{-5}\;GeV^{-2}$ as dictated by experimental
bounds only
the $\lambda'$'s.  Similarly the sensitivity of the invisible $\ups$ decay
to
the $Z'$ mass is given in Fig.~1b where now $X=M^2_Z/M^2_{Z'}$.

We next consider another unconventional contribution
to $\Gamma \;( \Upsilon, J/ \Psi \rightarrow \nu \bar \nu) $
- an extra neutral gauge boson, $Z'$, which provides an additional
annihilation
channel.  $Z'$ bosons appear as
remnants of a higher symmetry at large energies and have to be
sufficiently massive in order to fit current experimental
limits.
We will concentrate mainly on the superstring-inspired
$E_6$ grand unification model [10,7] and
left-right symmetric models [8].
In the models under consideration, the $Z'$ will correspond to
   an extra U(1) (for $E_6$) or to a neutral component of $SU(2)_{R}$
(for left-right models).

The phenomenology of an extra $Z'$ is highly model dependent [9].
For the case of $E_6$ this depends on the breaking of $E_6$ to the SM group.
The details are beyond the scope of this paper.  We are only interested in
probing $Z'$ with Eq. (1) and (2).  To this end, we first note that
we can neglect the $Z-Z'$ mixing because precision measurements
at LEP2 put a bound of $<0.0025$ for such mixings [9]. The fermion
couplings and hypercharges are uniquely determined by the way the Standard
Model is embedded in $E_6$ [10].
The neutral current process mediated by the $Z'$ boson involving the
fermion $f$ is given by
\begin{eqnarray}
{\cal L}_{NC} = g_E (Y'_{fL}\overline{f}_L \gamma^\mu f_{L} + Y'_{fR}
\overline{f}_R \gamma^\mu f_R) Z'_\mu \, \, .
\end{eqnarray}
where $g_E = \sqrt{\frac{5}{3}} g \tan \theta_w$, $g$ is the SU(2) gauge
coupling, and $Y'_f$'s are the hypercharge of the fermions of a given
chirality.  The relevant charges for the superstring-inspired $E_6$
are: $Y'_{c_{L},b_{L}}=\frac{1}{3}
\sqrt{\frac{3}{5}}$, $Y'_{c_{R}} = -\frac{1}{3} \sqrt{\frac{3}{5}}$,
$Y'_{\nu}=- \frac{1}{6} \sqrt{\frac{3}{5}}$ and $Y'_{b_{R}} = \frac{1}{6}
\sqrt{\frac{3}{5}}$.
The result is
\begin{eqnarray}
&& \frac{\Gamma \;( \Upsilon \rightarrow \nu \bar \nu) }
         {\Gamma \;( \Upsilon \rightarrow e^{+} e^{-} )}
         \Bigg \vert_{\rm SM + SUSY + E_6} =
   \frac{9 G^2 M_{\Upsilon}^4}{64 \pi^2 \alpha^2}
   \Bigl( \;
   2 \Omega^2 + \nonumber\\
&& \Bigl[ \Omega +
   2 \Bigl( \frac{M_{Z}^2}{m_{\tilde b_{R}}^2} +
   \frac{M_{Z}^2}{m_{\tilde b_{L}}^2} \Bigr)
   \;\frac {\cos^2 \theta_{W}}{g^2} \sum_{i=1}^{3}\vert
\lambda'_{i33}\vert^2
   \Bigr]^2
   \Bigr) \;,
\end{eqnarray}
where $\Omega = -1 + \frac{4}{3} \sin^2\theta_w + \frac{4 \cos^2\theta_w
M^2_{Z}}{M^2_{Z'}} \frac{g^2_E}{g^2} Y'_\nu (Y'_{b_{L}} + Y'_{b_{R}})$.
   %\Bigl( 4- \frac{M_{Z}^2}{M_{Z'}^2} \Bigr) \sin^2 \theta_{W}$.
Interestingly, the SM result for $J/\Psi$ width
is unaffected by the extra $E_6$-$Z'$ boson since
$Y'_{c_{L}} + Y'_{c_{R}} = 0 $. In contrast with R-breaking processes,
the $Z'$ increases the $\Upsilon$ width.  Hence, if both of these new
physics sources are present,  destructive interference
between these non-SM corrections can take place if the parameters are
favorable.
This would be fortuitous though not impossible.

For the case of the left-right model, both (1) and (2) undergo a certain
modification. Assuming $\frac{M_{Z}^2}{M_{Z'}^2}  \ll 1$,
the mixing angle $\phi$ between $Z$ and $Z'$ can be expressed as [8]
\begin{eqnarray}
\phi \simeq \sqrt{\cos
2\theta_{W}}\;\frac{M_{Z}^2}{M_{Z'}^2}\;.
\end{eqnarray}
Then
\begin{eqnarray}
&& \frac{\Gamma (J/ \Psi \rightarrow \nu  \bar \nu) )}
         {\Gamma (J/ \Psi \rightarrow e^{+} e^{-} )}
         \Bigg \vert_{\rm SM + SUSY + L-R}=
         \frac {27 G^2 M^4_{J/ \Psi} }{256 \pi^2 \alpha^2}
         (1 - \frac{8}{3} \sin^2 \theta_{W} )^2 \Delta^2  \;,
\end{eqnarray}
where
\begin{eqnarray}
\Delta=1-\Bigl( 1 - \frac{2 \sin^4
\theta_{W}}{\cos 2\theta_{W}}
\Bigr) \frac{M_{Z}^2}{M_{Z'}^2}\;. \nonumber
\end{eqnarray}
One readily obtains the corresponding expression for the $\Upsilon$ 
decay width from (8)
with
\begin{eqnarray}
\Omega \Rightarrow (-1 + \frac{4}{3} \sin^2 \theta_{W}) \Delta \;.
\end{eqnarray}
In this case the widths for both quarkonia will  decrease as compared to
their SM values.

In Fig.~2
we show the sensitivity
of the $\ups$ decay to the extra Z boson in the left-right symmetric model
with $x({\rm{LR}}) = M^2_Z/M^2_{Z'}$. Very similar behaviour for the branching
ratio of J/$\psi$ decay is displayed in Fig. 2b.  Evidently,
the latter resonance will not be suitable for probing this class of
models.

We have argued that because of the theorectically clean nature of the
decays a careful measurement of the invisible widths of the heavy
quarkonium states
can therefore yield constraints on a variety of physics beyond the
Standard
Model. It is especially useful for studying R-parity breaking terms of the
third generation in SUSY models. It is also sensitive to extra Z bosons
of GUTs or the left-right symmetric model. We also showed that the
 invisible channel will have to be due to the light neutrinos since the
other possibility of light gravitinos will not be significant. Neutralinos
will also not contribute since a lower bound of 40 GeV has already been
established by LEP measurements [11]. If a deviation from the SM value is
found in the $\ups$
decay a similar measurement for the J/$\psi$ can shed light on the source
of new physics.

The work is supported in part by the US Department of Energy under 
Grant No.\ DE-FG05-92ER40709-A005,  and the Natural Science and
Research Council of Canada.

\newpage

\section{Figure Captions}

\begin{enumerate}
\item[Fig. 1a]  The branching ratio of $\Upsilon \to \nu
    \overline{\nu}/\Upsilon \to e^+ e^-$ as a function of the parameter
    $x(SUSY)~=~{\sum_{i}} \mid \lambda_{i33}'\mid^2/m^2_{\tilde{b}}$ in
R-parity
    violating MSSM.  $\lambda'$ and the $m_{\tilde b}$ are given in the
text.
\item[Fig. 1b]  The branching ratio $\Upsilon \to \nu \overline{\nu} /
\Upsilon
  \to e^+ e^-$ as a function of $x \equiv M^2_{Z}/M^2_{Z^{\prime}} $ for
the extra
  $Z'$-boson in E(6) models.
\item[Fig. 2a]  The branching ratio $\Upsilon \to \nu
\overline{\nu}/\Upsilon
  \to e^+e^-$ as a function of $x(LR) \equiv {M^2_Z/M_Z^\prime}^{2}$ in
the
  left-right symmetric model.
\item[Fig. 2b]  The branching ratio of $J/\Psi \to \nu \overline{\nu}/\;
J/\Psi
      \to e^+e^-$  as a function of $x(LR)$.
\end{enumerate}

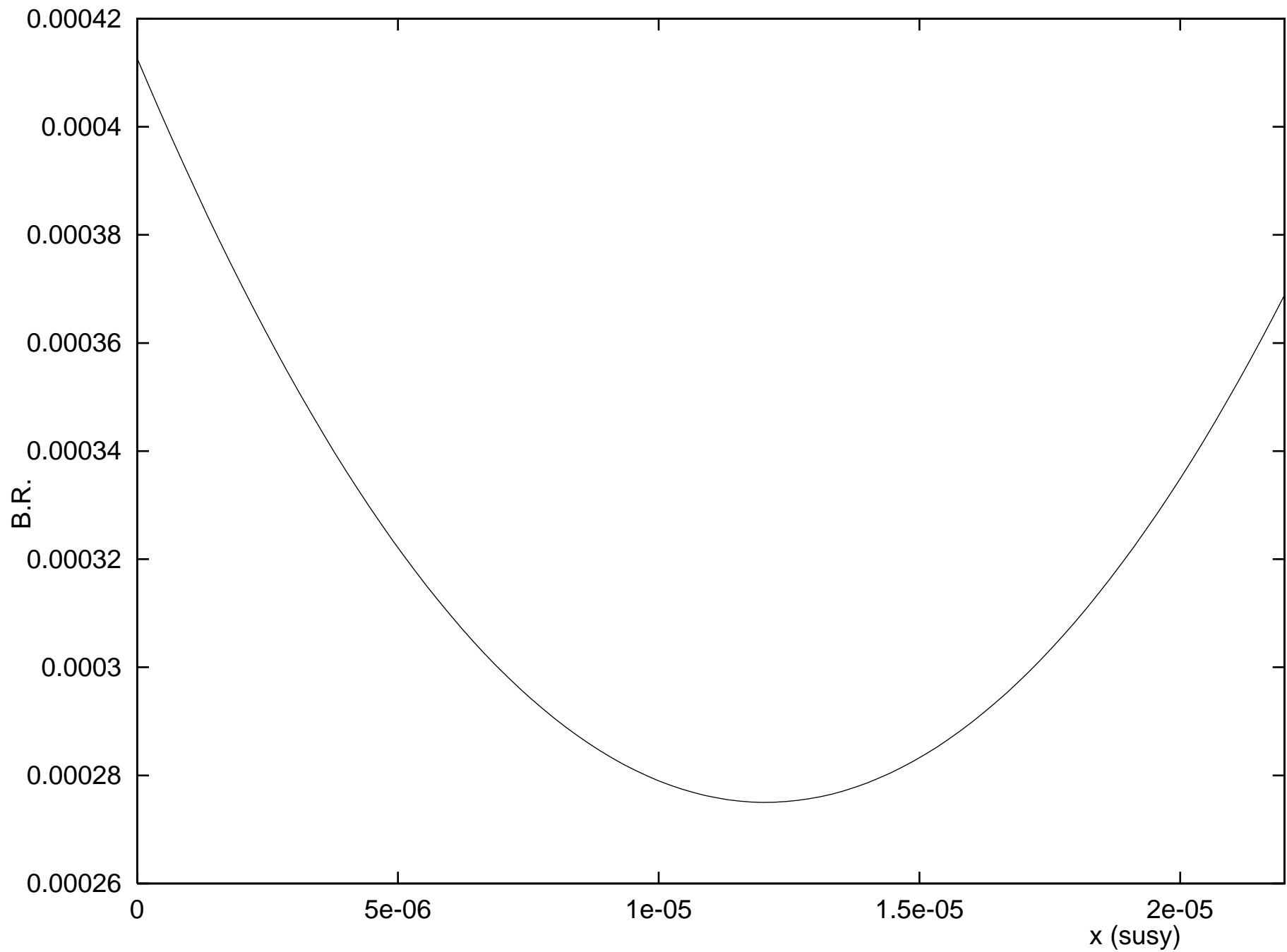
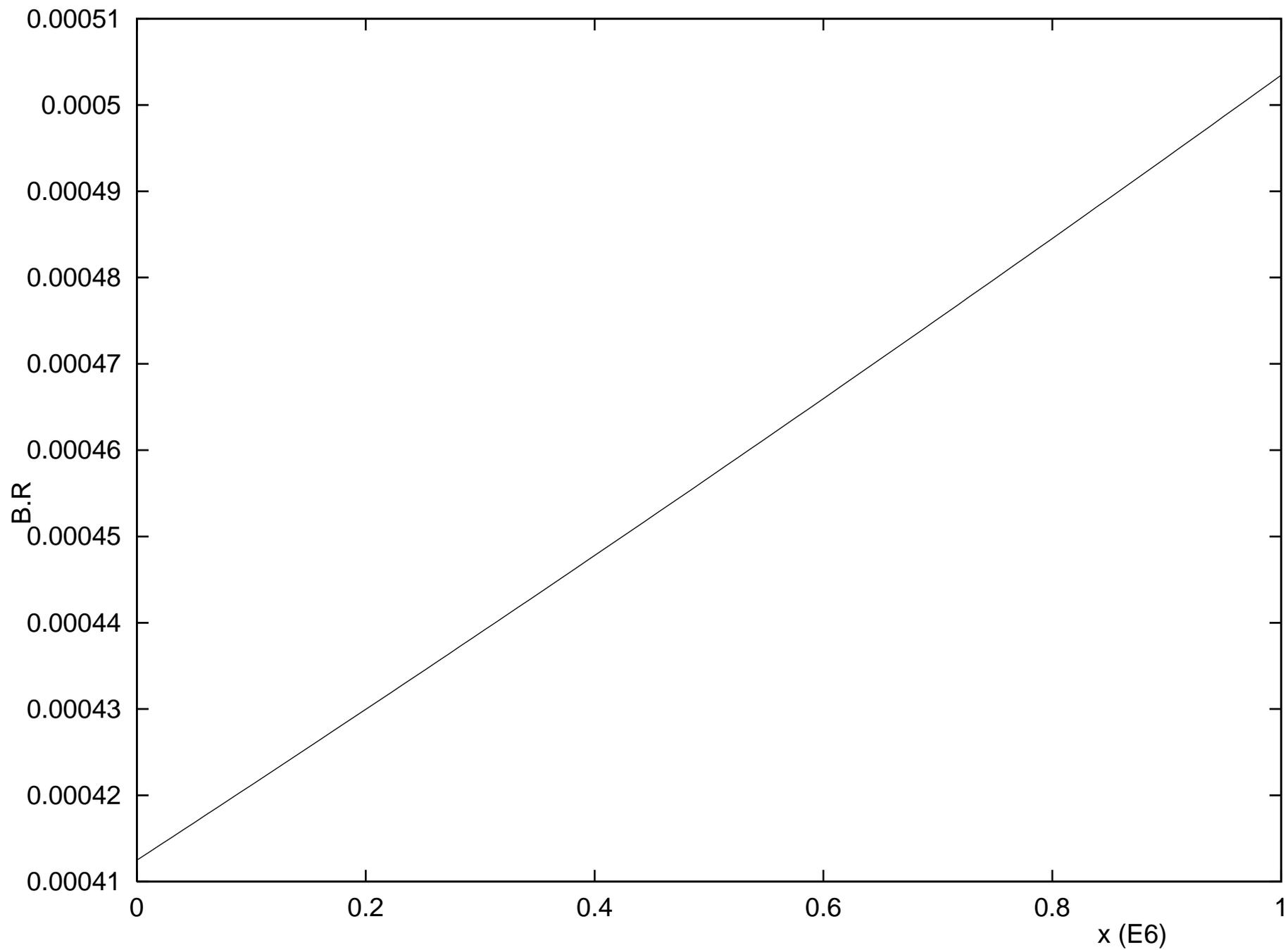
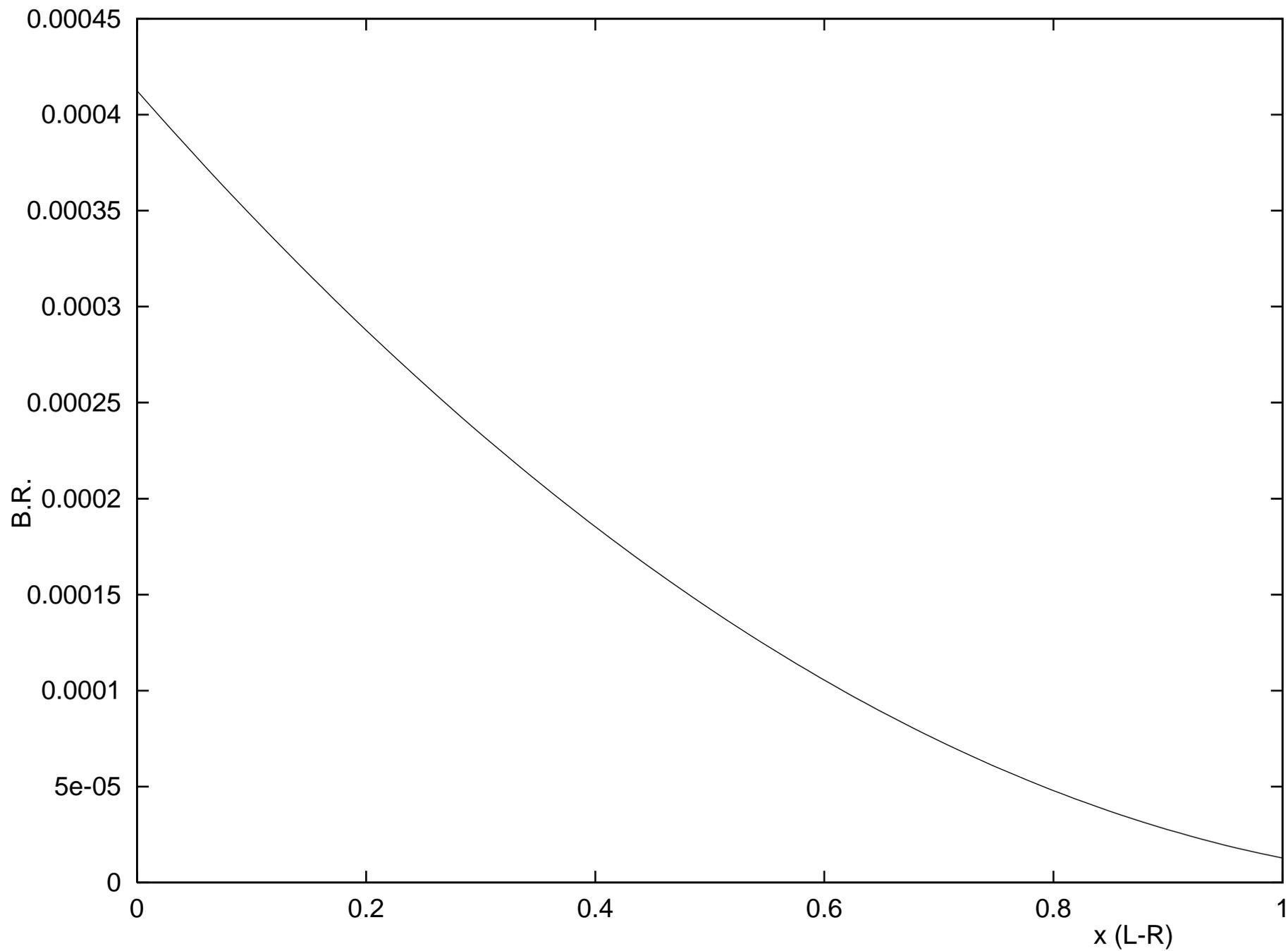
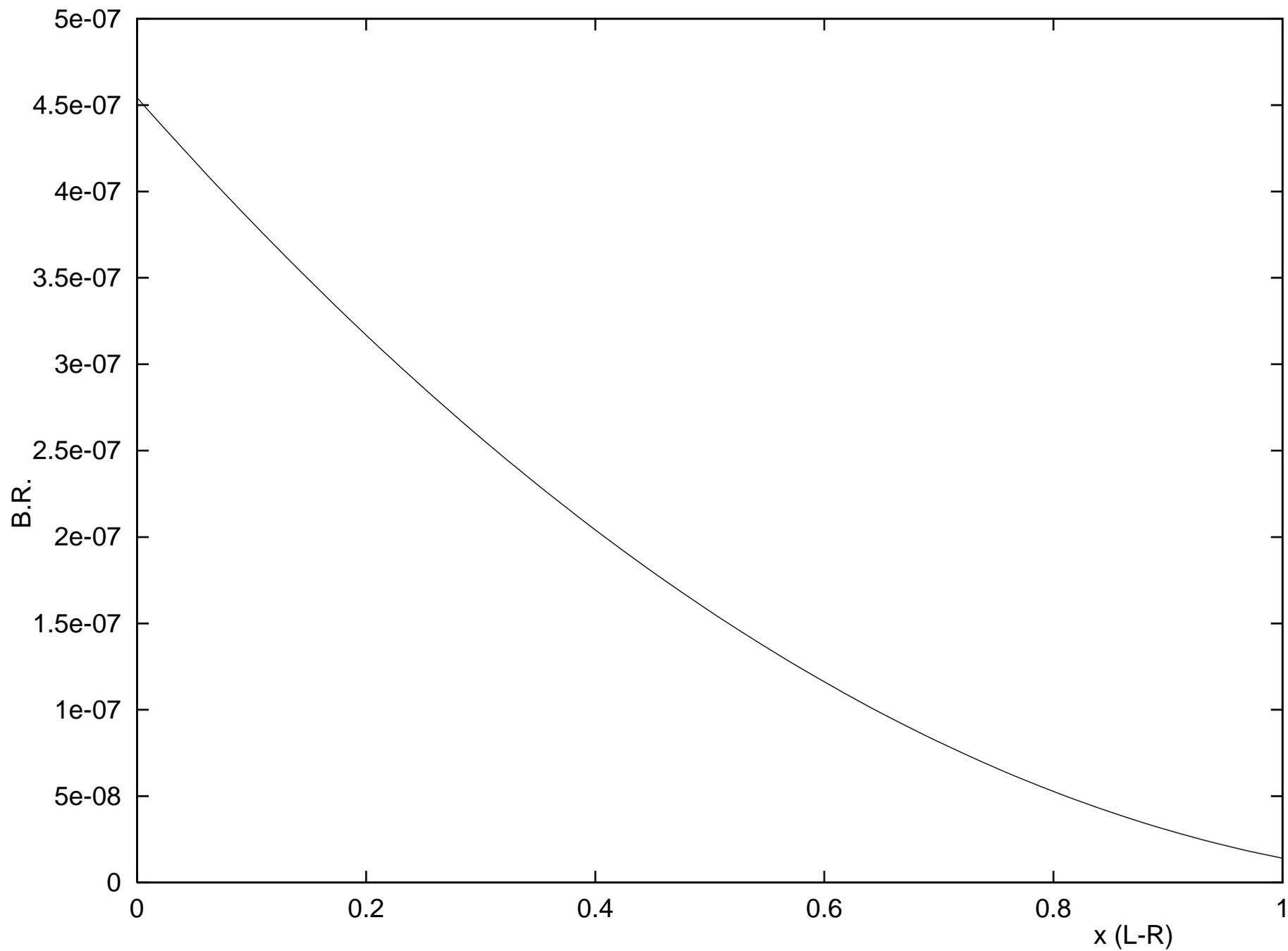
\end{document}